\documentclass[epj]{webofc}
\usepackage[utf8]{inputenc}
\usepackage[varg]{txfonts}   
\usepackage{booktabs}
\usepackage{xcolor}
\definecolor{darkred}{rgb}{0.4,0.0,0.0}
\definecolor{darkgreen}{rgb}{0.0,0.4,0.0}
\definecolor{darkblue}{rgb}{0.0,0.0,0.4}
\usepackage[bookmarks,linktocpage,colorlinks,
    linkcolor = darkred,
    urlcolor  = darkblue,
    citecolor = darkgreen]{hyperref}
\usepackage[FIGBOTCAP,TABBOTCAP,tight]{subfigure}
\wocname{EPJ Web of Conferences}
\woctitle{Lattice2017}
\usepackage{amsmath}
\usepackage{amssymb}
\usepackage{verbatim}
\usepackage{graphicx}
\usepackage{array}
\usepackage{multirow}
\usepackage{caption}
\usepackage{mdframed}
\usepackage{color}
\usepackage{listings}
\usepackage{bbold}
\usepackage[capitalise]{cleveref}

\graphicspath{{./figs/}}

\newcommand{\Tr}{\mathrm{Tr}\;}

\begin{document}
%
\selectlanguage{english}
\title{ How to identify zero modes for improved staggered fermions }
\subtitle{ Chiral Ward identities for Dirac eigenmodes with
  staggered fermions }
\author{%
  \firstname{Hwancheol} \lastname{Jeong} \inst{1} \fnsep\thanks{Speaker, \email{sonchac@gmail.com}} \and
  \firstname{Seungyeob} \lastname{Jwa} \inst{1} \and
  \firstname{Jangho} \lastname{Kim} \inst{2,3} \and
  \firstname{Sunghee} \lastname{Kim} \inst{1} \and
  \firstname{Sunkyu} \lastname{Lee} \inst{1} \and
  \firstname{Weonjong} \lastname{Lee}\inst{1}
  \fnsep\thanks{\email{wlee@snu.ac.kr}} \and
  \firstname{Jeonghwan} \lastname{Pak} \inst{1} \and
  \lastname{(SWME Collaboration)}
}
\institute{%
  Lattice Gauge Theory Research Center, CTP,
  Department of Physics and Astronomy, \\
  Seoul National University, Seoul 08826, South Korea
  \and
  National Superconducting Cyclotron Laboratory, Michigan State University,
  East Lansing, Michigan 48824, USA
  \and
  Department of Physics and Astronomy, Michigan State University, East
  Lansing, Michigan 48824, USA
}
\abstract{ We present results of the eigenvalue spectrum for the
  staggered Dir\"ac operator obtained using a modified Lanczos
  algorithm. We identify zero modes and non-zero modes.  We derive the
  chiral Ward identity derived from the conserved $U(1)_A$ symmetry,
  and check it numerically. This is the first step toward construction
  of an improved method to identify zero modes reliably with staggered
  fermions. }
\maketitle

\section{Introduction}
\label{sec:intro}

It is important to identify zero modes for any physical observable
sensitive to the topological charge such as quark condensates and
chiral condensates \cite{ Smit:1986fn, Follana:2006kb, Follana:2004sz,
  Follana:2005km, Durr:2004as, Donald:2011if, Cundy:2016tmw}.
Here, we present the first step toward an improved method to identify
zero modes using the chirality measurement and the chiral Ward
identities derived from the conserved $U(1)_A$ symmetry in staggered
fermions.
%

\section{Notations and definitions}
\label{sec:def}

Here, we adopt the notation of Ref.~\cite{Lee:2001hc} except for the
gauge links.
In general, a staggered bilinear operator is defined as
\begin{align}
  \mathcal{O}_{S\times T} (x) &\equiv \overline{\chi} (x_A)\, [\gamma_S
    \otimes \xi_T ]_{AB}\, \chi(x_B) \nonumber\\
  &= \overline{\chi}_a (x_A)\, \overline{(\gamma_S \otimes \xi_T)}_{AB}\,
  U(x_A,x_B)_{ab}\, \chi_{b}(x_B) \,.
\end{align}
Here $\chi$ is the staggered fermion field, $a,b$ are color indices, and
$x_A = 2x + A$ where $A,B$ are hypercubic vectors with $A_\mu, B_\mu \in
\{0,1\}$.
\begin{equation}
  \overline{(\gamma_S \otimes \xi_T)}_{AB} = \frac{1}{4} \Tr
  (\gamma_A^\dagger \gamma_S \gamma_B \gamma_T^\dagger) \,,
\end{equation}
where $\gamma_S$ represents Dirac spin matrix, and $\xi_T$ represents the
$4\times 4$ taste matrix.
\begin{equation}
  U(x_A,x_B) \equiv \mathbb{P}_{\mathrm{SU(3)}} \left[
    \sum_{p\in\mathcal{C}} V(x_A,x_{p_1}) V(x_{p_1}, x_{p_2}) \cdots
    V(x_{p_n}, x_B) \right] \,,
\end{equation}
where $\mathbb{P}_{\mathrm{SU(3)}}$ is the SU(3) projection, $\mathcal{C}$
represents a complete set of the shortest paths from $x_A$ to $x_B$, and
$V(x,y)$ represents HYP-smeared fat links
\cite{Hasenfratz:2001hp,Lee:2002ui} for HYP staggered fermions
\cite{Hasenfratz:2001hp}, Fat7 fat links
\cite{Lee:2002ui,Lee:2002fj,Orginos:1999cr,Lepage:1998vj} for asqtad
\cite{Bazavov:2009bb} or HISQ \cite{Follana:2006rc} staggered fermions, and
thin gauge links for unimproved staggered fermions.
%

\section{Eigenmodes of staggered fermions}
\label{sec:eigenmodes}

Let $D_s$ be Dirac operator for staggered fermions.
Here we call \emph{staggered fermions} collectively including unimproved
staggered fermions and various improved staggered fermions such as
HYP-smeared staggered fermions \cite{Hasenfratz:2001hp}, asqtad staggered
fermions \cite{Bazavov:2009bb}, and HISQ staggered fermions
\cite{Follana:2006rc}.
Regardless of improvement methods, staggered Dirac operator $D_s$ is
anti-Hermitian: $D_s^\dagger = - D_s$.
This restricts eigenvalues of staggered fermions to be purely imaginary or
zero:
\begin{equation}
  D_s | f_\lambda^s \rangle = i \lambda\, | f_\lambda^s \rangle \,,
\end{equation}
where $\lambda$ is real, and $| f_\lambda^s \rangle$ is an eigenvector
corresponding to the eigenvalue $i \lambda$.
The superscript $s$ represents the staggered fermions.
In the meantime, the staggered Dirac operator anti-commutes with the
operator $\Gamma_\varepsilon = [\gamma_5 \otimes \xi_5]$, the generator for
the $\mathrm{U(1)}_A$ symmetry: $\Gamma_\varepsilon D_s = - D_s\,
\Gamma_\varepsilon$.
For a given eigenvalue $i\lambda$ of $D_s$ and its corresponding
eigenvector $| f_\lambda^s \rangle$,
\begin{equation}
  \label{eq:negeig}
  D_s\, \Gamma_\varepsilon\, | f_\lambda^s \rangle = - i \lambda\,
  \Gamma_\varepsilon\, | f_\lambda^s \rangle \,.
\end{equation}
Hence there must exist another eigenvector $| f_{-\lambda}^s \rangle
\propto \Gamma_\varepsilon\, | f_{+\lambda}^s \rangle$ with
the eigenvalue $-i\lambda$.
This insures that eigenvalues of staggered fermions must exist as
$\pm$ pair except for zero modes.
In practice, we calculate eigenvalues and eigenvectors of $D_s^\dagger
D_s$ instead of $D_s$.
The $D_s^\dagger D_s$ is a Hermitian and positive semi-definite
operator.
The eigenvalue equation of $D_s^\dagger D_s$ can be written by
\begin{equation}
  D_s^\dagger D_s\, |\, g_{\lambda^2}^s \rangle = \lambda^2\, |\,
  g_{\lambda^2}^s \rangle \,.
\end{equation}
Eigenvalues of a Hermitian operator can be calculated faster than
those of non-Hermitian operators.
A positive semi-definite operator has non-negative eigenvalues such
that we can restrict the range of eigenvalues to the non-negative
region.
In addition, the operator $D_s^\dagger D_s$ allows us to treat even
sites and odd sites separately by even-odd preconditioning \cite{
  DeGrand:1990dk}.

For each eigenvalue $\lambda^2$ of $D_s^\dagger D_s$, there exist
two corresponding eigenvalues $\pm i\lambda$ of $D_s$.
Hence, $|\, g_{\lambda^2}^s\rangle$ is composed of both
$| f_{+\lambda}^s \rangle$ and $| f_{-\lambda}^s \rangle$.
In other words,
\begin{equation}
  |\, g_{\lambda^2}^s \rangle = c_1\, | f_{+\lambda}^s \rangle + c_2\, |
  f_{-\lambda}^s \rangle \,,
\end{equation}
where $c_1$ and $c_2$ are complex numbers satisfying the normalization
condition $|c_1|^2+|c_2|^2=1$.
In fact, we can obtain the eigenvectors $| f_{+\lambda}^s \rangle$ and
$| f_{-\lambda}^s \rangle$ from $|\, g_{\lambda^2}^s \rangle$ by
projection.
Let us consider projection operators given by
\begin{align}
  P_\pm \equiv (D_s \pm i\lambda) \,.
\end{align}
Note that $P_+$ removes $| f_{-\lambda}^s \rangle$ component, and $P_-$
removes $|f_{+\lambda}^s \rangle$ component.
Applying them to $|\, g_{\lambda^2}^s \rangle$,
\begin{align}
  |\, \psi_+ \rangle &\equiv P_+ |\, g_{\lambda^2}^s \rangle = c_1\, 2i\lambda\, |
  f_{+\lambda}^s \rangle \,,\\
  |\, \psi_- \rangle &\equiv P_- |\, g_{\lambda^2}^s \rangle = - c_2\, 2i\lambda\, |
  f_{-\lambda}^s \rangle \,.
\end{align}
Now normalizing $|\, \psi_+ \rangle$ and $|\, \psi_- \rangle$ give the
eigenvectors $| f_{+\lambda}^s \rangle$ and $| f_{-\lambda}^s \rangle$:
\begin{align}
  | f_{\pm\lambda}^s \rangle &= \frac{ |\, \psi_\pm \rangle }{ \sqrt{ \langle
      \psi_\pm | \psi_\pm \rangle } } \,.
  \label{eq:f_pm}
\end{align}
%

\section{Eigenvalue spectrum with HYP staggered fermions}
\label{sec:eigenvalues}

We use a variation of Lanczos algorithm to calculate eigenvalues and
eigenvectors \cite{Lanczos:1950zz}.
It is adapted to our purpose with several improvement techniques, such as
the implicit restart \cite{Lehoucq96deflationtechniques} and the polynomial
acceleration with Chebyshev polynomial \cite{MR736453}.
We are interested only in small eigenvalues near zero.
However, it is possible but quite slow if we try to make the
eigenvalues of submatrices converge to those small eigenvalues using
the typical Lanczos procedure.
Hence, we need to introduce an acceleration method to speed up the
convergence of Lanczos.
We adopt the Chebyshev polynomial acceleration method \cite{MR736453} here.
\begin{table}[htbp]
  \small
  \centering
  \sidecaption
  \begin{tabular}[b]{r|l}
    \hline\hline
    parameter & value \\
    \hline
    gluon action & tree level Symanzik \cite{Luscher:1984xn,Luscher:1985zq,Alford:1995hw} \\
    tadpole improvement & yes \\
    $\beta$ & 4.6 \\
    geometry & $12^4$ \\
    $a$ & 0.125 fm \\
    \hline
    valence quarks & HYP staggered fermions \cite{Lee:2002ui,Kim:2010fj,Kim:2011pz} \\
    $N_f$ & $N_f=0$ (quenched QCD) \\
    \hline\hline
  \end{tabular}
  \caption{Simulation environment}
  \label{tab:para}
\end{table}
Here, we perform our numerical study in quenched QCD in this paper.
Details on our simulation and the gauge ensemble are described in
Table~\ref{tab:para}.
We use HYP staggered fermions \cite{Hasenfratz:2001hp} as valence quarks in our
numerical study.
\begin{figure}[tbhp]
  \centering
  \subfigure[$\lambda_i^2$]{
    \includegraphics[width=0.48\linewidth]{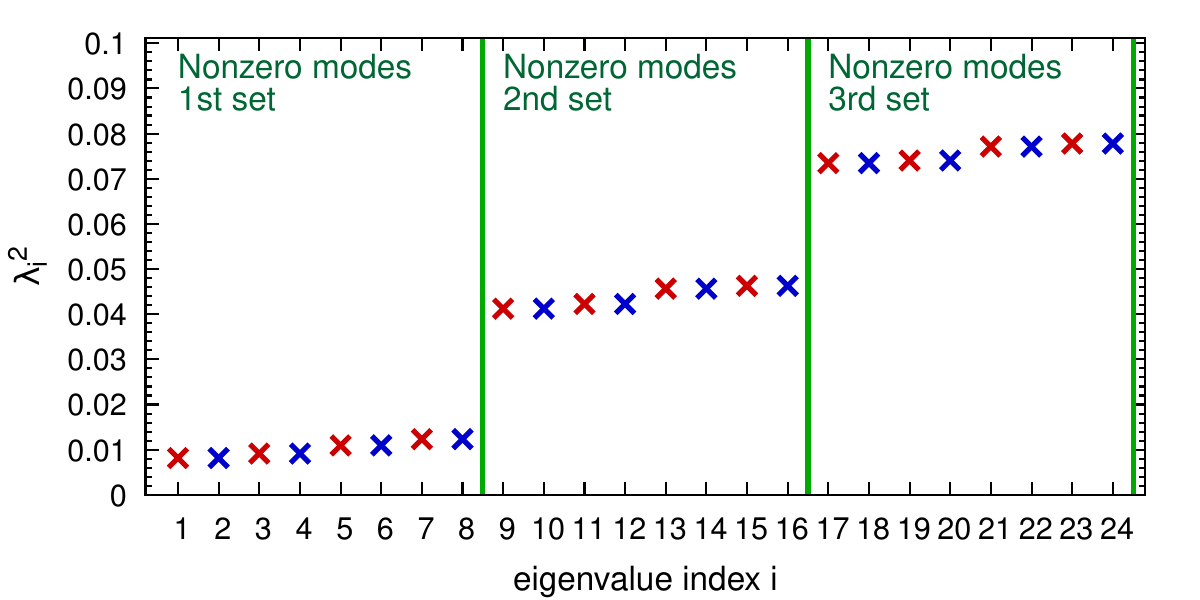}
    \label{fig:q0_abs}
  }
  \subfigure[$\lambda_i$]{
    \includegraphics[width=0.48\linewidth]{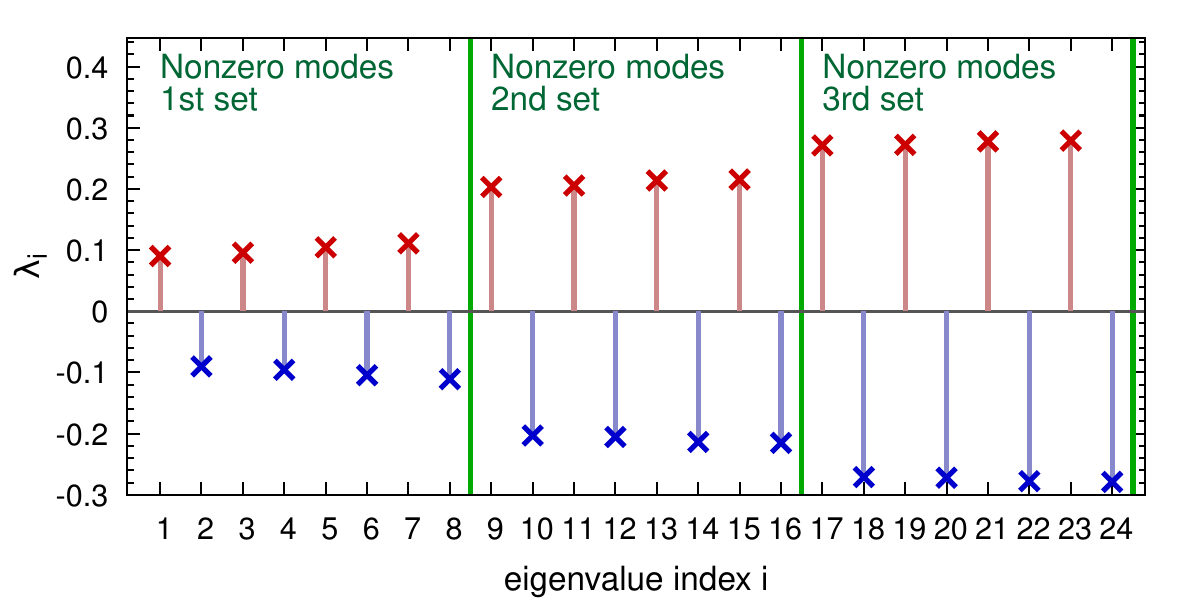}
    \label{fig:q0_ev}
  }
  \caption{ Eigenvalue spectrum of staggered Dirac operator on a gauge
    configuration with $Q=0$. Eigenvalues are sorted by their values
    in ascending order.}
  \label{fig:ev_q0}
\end{figure}
In Fig.~\ref{fig:ev_q0}, we present the eigenvalue spectrum on a gauge
configuration with topological charge $Q=0$.
Topological charge $Q$ is determined using gauge links through the APE
smearing \cite{Albanese:1987ds}.
In Fig.~\ref{fig:ev_q0}\;\subref{fig:q0_abs}, we present results of
small eigenvalues $\lambda_i^2$ of the $D_s^\dagger D_s$ operator.
In Fig.~\ref{fig:ev_q0}\;\subref{fig:q0_ev}, we present results of
small eigenvalues $\lambda_i$ of the $D_s$ operator.
Here, recall that the Lanczos algorithm produces $\lambda_i^2$
eigenvalues of $D_s^\dagger D_s$ and eigenvectors $|\,
g_{\lambda_i^2}^s \rangle$.
Using Eq.~\eqref{eq:f_pm}, we can obtain $\pm\lambda_i$ eigenvalues of
$D_s$ and eigenvectors $|\, f_{\pm\lambda_i}^s \rangle$.
In Fig.~\ref{fig:ev_q0}\;\subref{fig:q0_ev}, we assign indices of
eigenvalues such that $\lambda_{2n} = - \lambda_{2n-1}$
for $n>0$ with $n \in \mathbb{Z}$.
For each non-zero eigenvalue $\lambda_i$, we find the four-fold
degeneracy near $\lambda_i$ due to the approximate SU(4) taste
symmetry, and for each of them there exists a $U(1)_A$ parity partner
eigenmode with $\lambda = -\lambda_i$.
Hence, we find an eight-fold degeneracy for each non-zero eigenmode as
in Ref.~\cite{Follana:2004sz, Follana:2005km, Follana:2006kb,
  Durr:2004as}.

\begin{figure}[tbhp]
  \centering
  \subfigure[$\lambda_i^2$]{
    \includegraphics[width=0.48\linewidth]{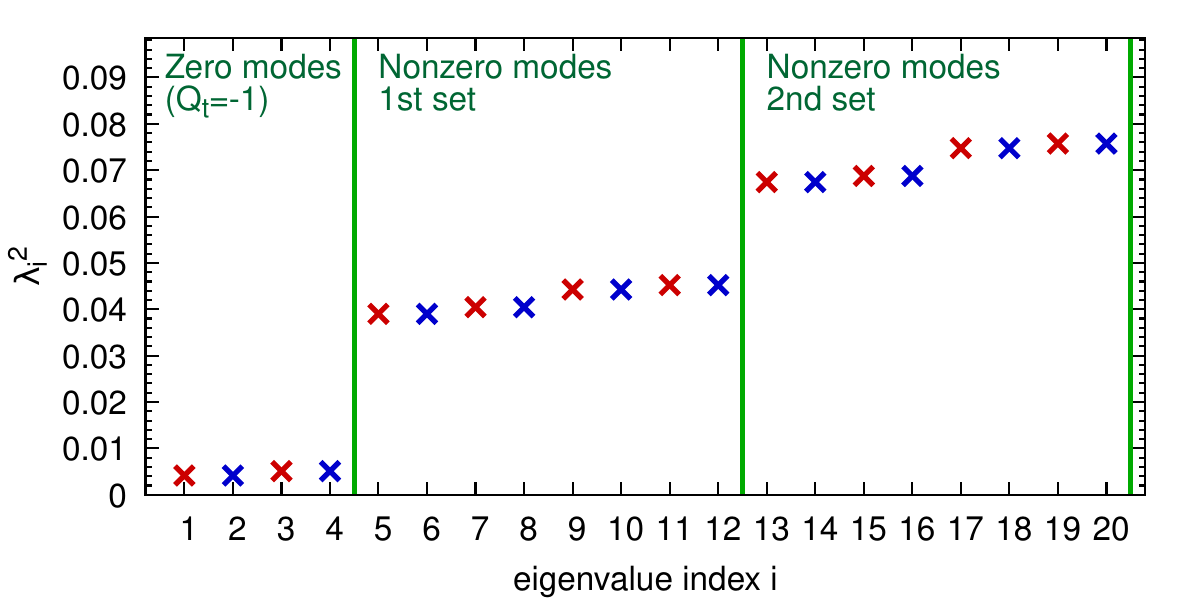}
    \label{fig:qm1_abs}
  }
  \subfigure[$\lambda_i$]{
    \includegraphics[width=0.48\linewidth]{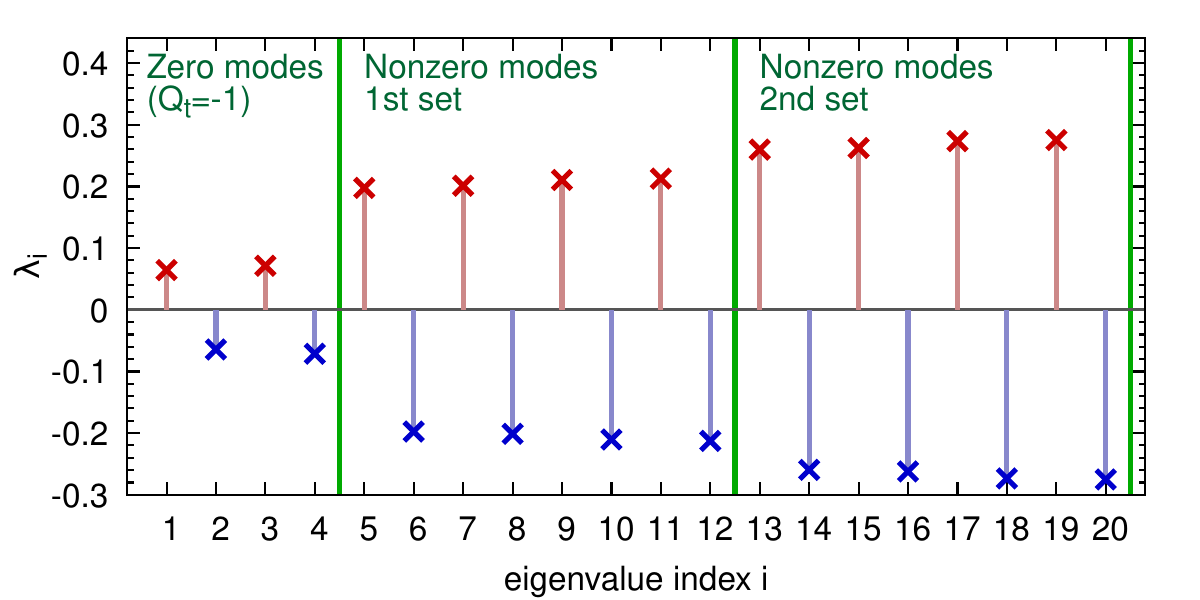}
    \label{fig:qm1_ev}
  }
  \caption{The same as Fig.~\ref{fig:ev_q0} except $Q=-1$.}
  \label{fig:ev_qm1}
\end{figure}
In Fig.~\ref{fig:ev_qm1}, we present the eigenvalue spectrum on a gauge
configuration with topological charge $Q=-1$.
In the previous example with $Q=0$, there is no zero mode.
In case of $Q=-1$, we expect to see four-fold degenerate zero modes
thanks to the approximate SU(4) taste symmetry in staggered fermions
as in Refs.~\cite{Follana:2004sz, Follana:2005km, Follana:2006kb,
  Durr:2004as}.
In Fig.~\ref{fig:ev_qm1}, the first four eigenvalues corresponds to
the would-be four-fold degenerate zero modes.
They show up in pairs due to the conserved $U(1)_A$ symmetry.
In other words, they satisfy $\lambda_{2n} = - \lambda_{2n-1}$:
$\lambda_2 = - \lambda_1$ and $\lambda_4 = - \lambda_3$.
For more details, refer to \cite{prd:prep}.
In principle, one can distinguish would-be zero modes from non-zero
modes by counting the number of degenerate states: four-fold for zero
modes and eight-fold for non-zero modes.
However, this idea might not work well when the eigenvalue spectrum is
too dense in a large volume to classify them into groups of degenerate
eigenvalues.
Hence, it is necessary to have a significantly improved method to
identify zero modes more reliably.
We will address this issue in following sections.
%

\section{Chiral symmetry of staggered fermions}
\label{sec:chiral}

Let us consider the $U(1)_A$ operator $\Gamma_\varepsilon$ defined
in Sec.~\ref{sec:eigenmodes}.
Recall that an eigenmode $| f_{+\lambda}^s \rangle$ and its parity
partner state $| f_{-\lambda}^s \rangle$ are related as in
Eq.~\eqref{eq:negeig}.
Hence, we find that
\begin{align}
  \label{eq:fp_to_fm}
  \Gamma_\varepsilon\, | f_{+\lambda}^s \rangle &= e^{+i\theta}\, |
  f_{-\lambda}^s \rangle \,,\\
  \label{eq:fm_to_fp}
  \Gamma_\varepsilon\, | f_{-\lambda}^s \rangle &= e^{-i\theta}\, |
  f_{+\lambda}^s \rangle \,.
\end{align}
Here, there is no constraint on the phase $\theta$.
Hence, we expect it would be real and random.

\begin{figure}[tbhp]
  \centering
  \subfigure[zero modes]{
    \includegraphics[width=0.48\linewidth]{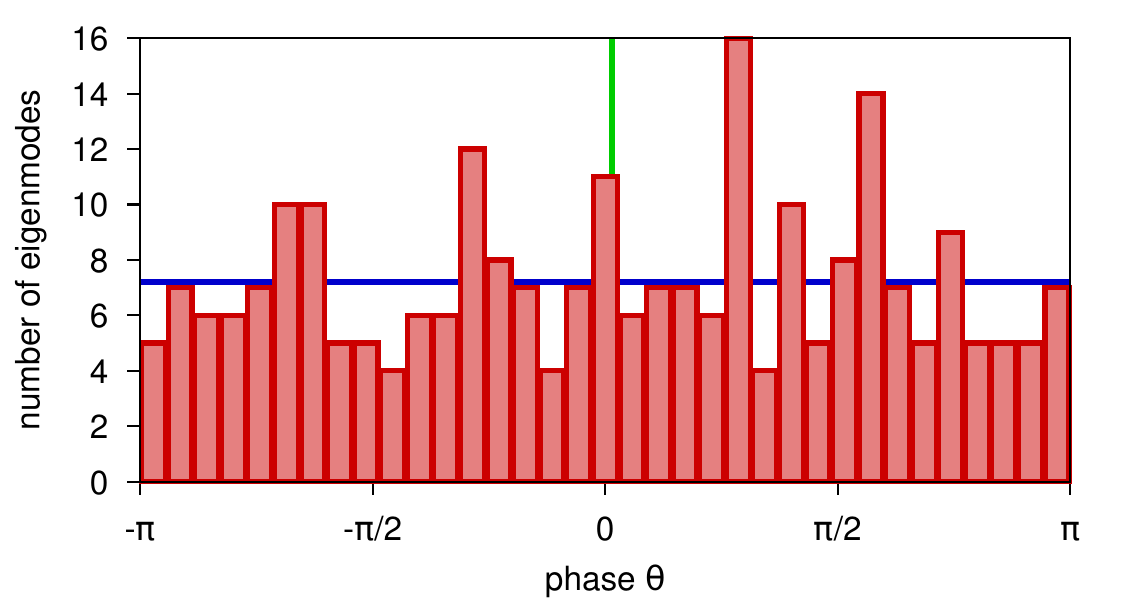}
    \label{fig:theta_zero}
  }
  \subfigure[non-zero modes]{
    \includegraphics[width=0.48\linewidth]{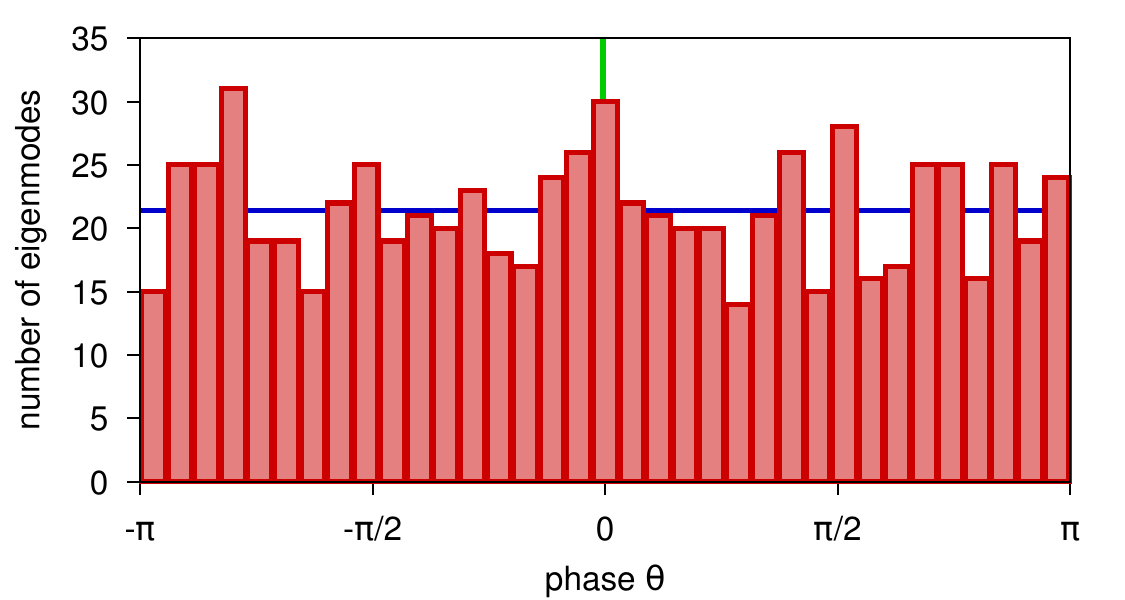}
    \label{fig:theta_nonzero}
  }
  \caption{Histograms of the phase for the Ward identity
    Eqs.~\eqref{eq:fp_to_fm} and \eqref{eq:fm_to_fp}}
  \label{fig:theta}
\end{figure}
In Fig.~\ref{fig:theta}, we present the probability distribution
of $\theta$ for both zero modes and non-zero modes.
We find that the phase $\theta$ is random in both cases.

Let us define the chirality $\Gamma_5$ as
\begin{equation}
  \Gamma_5 (\alpha,\beta) \equiv \langle f^s_\alpha | [\gamma_5 \otimes 1]
  | f^s_\beta \rangle \,,
\end{equation}
where $\alpha$ and $\beta$ represent eigenvalues of $D_s$.
We find that the lattice chirality operator $[\gamma_5 \otimes 1]$
satisfies the same recursion relations as the continuum chirality
operator $\gamma_5$:
\begin{gather}
  \label{eq:g5x1_r1}
  [\gamma_5 \otimes \mathbb{1}]^{2n+1} = [\gamma_5 \otimes \mathbb{1}] \,,
  \\
  \label{eq:g5x1_r2}
  [\gamma_5 \otimes \mathbb{1}]^{2n} = [\mathbb{1} \otimes \mathbb{1}] \,,
  \\
  \left[\frac{1}{2}(1\pm\gamma_5) \otimes \mathbb{1}\right]^n =
  \left[\frac{1}{2}(1\pm\gamma_5) \otimes \mathbb{1}\right] \,, \\
  \left[\frac{1}{2}(1+\gamma_5) \otimes \mathbb{1}\right] \left[\frac{1}{2}(1-\gamma_5)
    \otimes \mathbb{1}\right] = 0\,,
\end{gather}
where $n \geq 0$ and $n \in \mathbb{Z}$.
For more details on the rigorous proof, refer to Ref.~\cite{prd:prep}.
\begin{figure}[tbhp]
  \centering
  \subfigure[$Q=0$]{
    \includegraphics[width=0.48\linewidth]{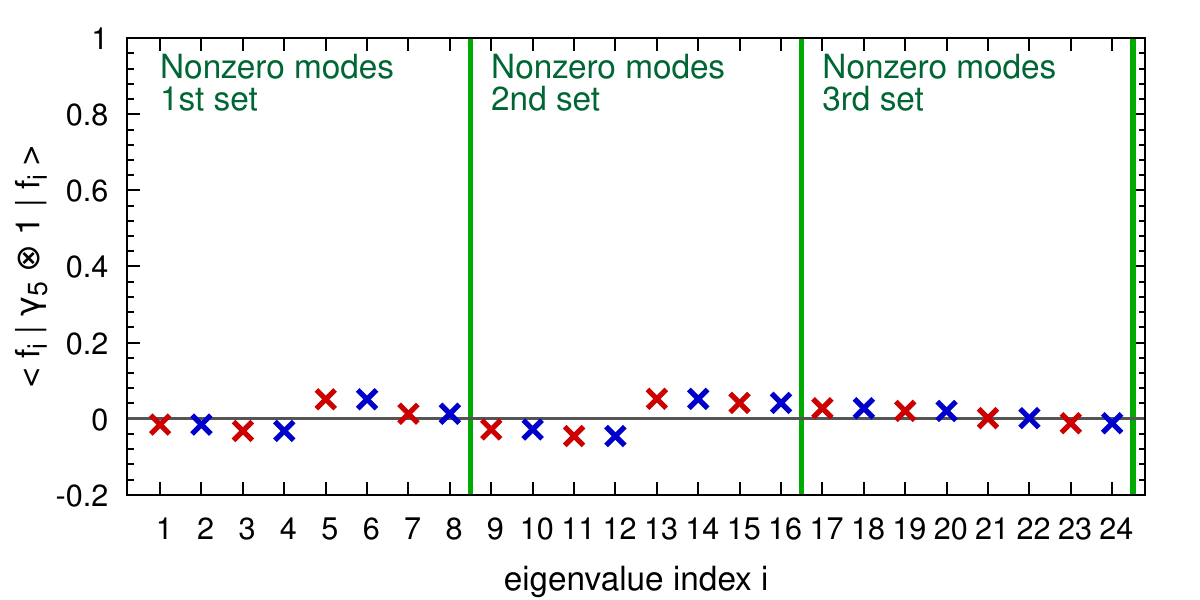}
    \label{fig:q0_g5}
  }
  \subfigure[$Q=-1$]{
    \includegraphics[width=0.48\linewidth]{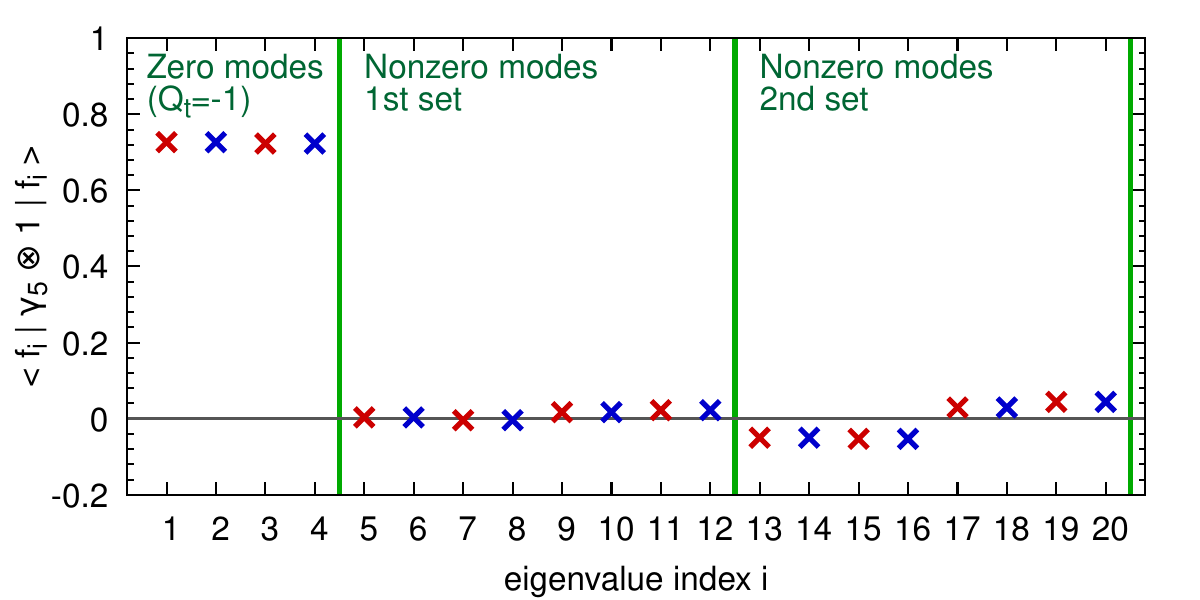}
    \label{fig:qm1_g5}
  }
  \subfigure[$Q=-2$]{
    \includegraphics[width=0.48\linewidth]{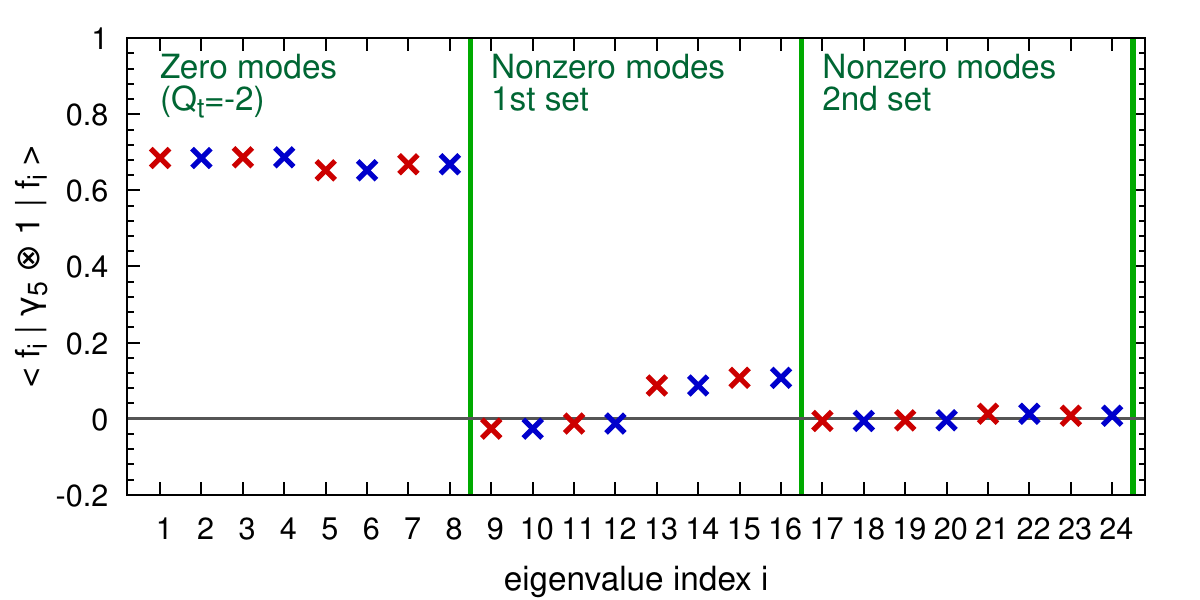}
    \label{fig:qm2_g5}
  }
  \subfigure[$Q=-3$]{
    \includegraphics[width=0.48\linewidth]{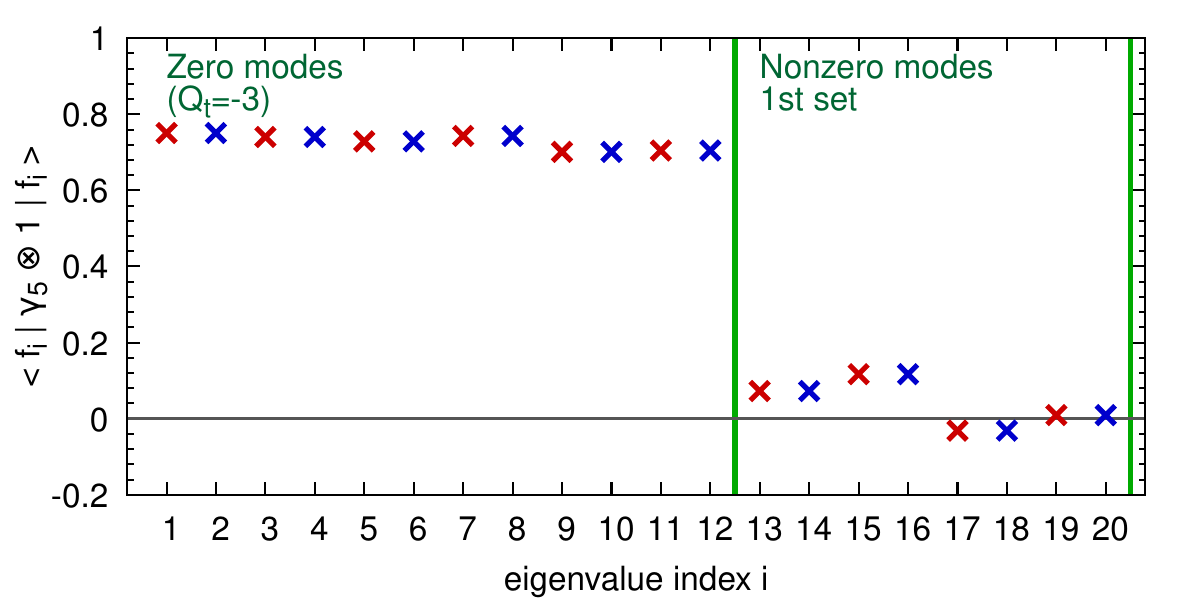}
    \label{fig:qm3_g5}
  }
  \caption{The chirality measurement of $\Gamma_5(\alpha,\alpha) =
    \langle f^s_\alpha | [\gamma_5 \otimes 1] | f^s_\alpha \rangle $
    for configurations with various topological charge $Q$. }
  \label{fig:g5}
\end{figure}

In Fig.~\ref{fig:g5}, we present results for the diagonal chirality
$\Gamma_5( \alpha, \alpha)$ for configurations with topological charge
$Q=0,-1,-2,-3$.
In the continuum, the diagonal chirality should be $\pm 1$ for zero modes
and 0 for non-zero modes.
On the lattice, the chiral symmetry for $[\gamma_5 \otimes 1]$
transformation is not conserved, and so the chirality operator
receives renormalization.
This causes the diagonal chirality to be around 0.7 instead of 1.
As one can see, the diagonal chirality measurement tells us how many zero
modes exist in a given configuration \cite{Follana:2004sz, Follana:2005km,
  Follana:2006kb}.
Hence, the diagonal chirality serves as a good criterion to distinguish
the zero modes from non-zero modes.
%

\section{Chiral Ward identity}
\label{sec:chiralWard}

Let us consider the \emph{shift operator} which is defined as:
\begin{equation}
  \Xi_5 (\alpha,\beta) \equiv \langle f^s_\alpha | [1 \otimes \xi_5] |
  f^s_\beta \rangle \,.
\end{equation}
Similar to the chirality operator $[\gamma_5 \otimes 1]$, the shift
operator satisfies the following recursion relations:
\begin{gather}
  \label{eq:1xxi5_r1}
  [ 1 \otimes \xi_5 ]^{2n+1} = [ 1 \otimes \xi_5 ] \,,\\
  \label{eq:1xxi5_r2}
  [ 1 \otimes \xi_5 ]^{2n} = [1 \otimes 1 ] \,, 
\end{gather}
for $n \geq 0$ and $n \in \mathbb{Z}$.
Using Eqs.~\eqref{eq:g5x1_r1}, \eqref{eq:g5x1_r2}, \eqref{eq:1xxi5_r1}, and
\eqref{eq:1xxi5_r2},
one can find that the following relations hold:
\begin{gather}
  \Gamma_\varepsilon\, [ \gamma_5 \otimes 1 ] = [ \gamma_5 \otimes 1 ]\,
  \Gamma_\varepsilon = [ 1 \otimes \xi_5 ] \,,\\
  \Gamma_\varepsilon\, [ 1 \otimes \xi_5 ] = [ 1 \otimes \xi_5 ]\,
  \Gamma_\varepsilon = [ \gamma_5 \otimes 1 ] \,.
\end{gather}
Applying them to eigenstates $| f_{\pm\lambda}^s \rangle$,
\begin{gather}
  \label{eq:chiWard1}
  e^{+i\theta} [ \gamma_5 \otimes 1 ]\, | f_{-\lambda}^s \rangle = [ 1
    \otimes \xi_5 ]\, | f_{+\lambda}^s \rangle \,,\\
  \label{eq:chiWard2}
  e^{-i\theta} [ \gamma_5 \otimes 1 ]\, | f_{+\lambda}^s \rangle = [ 1
    \otimes \xi_5 ]\, | f_{-\lambda}^s \rangle \,.
\end{gather}
Eqs.~\eqref{eq:chiWard1} and \eqref{eq:chiWard2} are Ward identities for
the chirality of staggered fermions.
%
%
We can go further.
Multiplying $\langle f_{\pm\alpha}^s |$ to the left and replacing the
notation $\lambda$ to $\beta$, we have
\begin{gather}
  e^{+i\theta} \Gamma_5 (\pm\alpha,-\beta) = \Xi_5 (\pm\alpha,+\beta) \,,\\
  e^{-i\theta} \Gamma_5 (\pm\alpha,+\beta) = \Xi_5 (\pm\alpha,-\beta) \,.
\end{gather}
Hence, we obtain the following simple Ward identities:
\begin{equation}
  \label{eq:chiWard3}
  |\,\Gamma_5 (\alpha,\beta)\,| = |\,\Xi_5 (-\alpha,\beta)\,| = |\,\Xi_5
  (\alpha,-\beta)\,| = |\,\Gamma_5 (-\alpha,-\beta)\,| \,.
\end{equation}
In addition, the Hermiticity provides additional Ward identities:
\begin{align}
  |\, \Gamma_5(\alpha,\beta)\, | &=|\, \Gamma_5(\beta,\alpha)\,| \,,
  \\
  |\, \Xi_5(\alpha,\beta)\, | &=|\, \Xi_5(\beta,\alpha)\,|  \,.
\end{align}
Then, we obtain the final form of the chiral Ward
identity:
\begin{align}
  \label{eq:chiWard4}
  |\,\Gamma_5 (\alpha,\beta)\,| &= |\,\Xi_5
  (-\alpha,\beta)\,| = |\,\Xi_5 (\alpha, -\beta)\,| = |\,\Gamma_5
  (-\alpha, -\beta)\,|
  \nonumber \\
  &= |\,\Gamma_5 (\beta, \alpha)\,| = |\,\Xi_5 (-\beta, \alpha)\,|
  = |\,\Xi_5 (\beta, -\alpha)\,| = |\,\Gamma_5 (-\beta, -\alpha)\,| \,.
\end{align}
\begin{table}[htbp]
  \small
  \caption{Numerical examples for the chiral Ward identity (WI). Here,
    we use the notation of $\lambda_{2n} = -\lambda_{2n-1}$ for $n \ge
    1$ and $n \in \mathbb{Z}$: for example, $\lambda_{12} =
    -\lambda_{11}$, and $\lambda_6 = -\lambda_5$.}
  \label{tab:WI}
  \renewcommand{\arraystretch}{1.2}
  \subtable[Diagonal WI]{
    \begin{tabular}[b]{ @{\qquad} c @{\qquad} c @{\qquad} }
      \hline\hline
      parameter & value \\
      \hline
      $|\,\Gamma_5 (\lambda_1,\lambda_1)\,|$ & 0.7268534 \\
      $|\,\Xi_5 (\lambda_2,\lambda_1)\,|$ & 0.7268534 \\
      $|\,\Xi_5 (\lambda_1,\lambda_2)\,|$ & 0.7268534 \\
      $|\,\Gamma_5 (\lambda_2,\lambda_2)\,|$ & 0.7268534 \\
      \hline\hline
    \end{tabular}
  }
  \hfill
  \subtable[Off-diagonal WI]{
    \begin{tabular}[b]{ @{\qquad} c @{\qquad} c @{\qquad} | @{\qquad} c @{\qquad} c @{\qquad} }
      \hline\hline
      parameter & value & parameter & value \\
      \hline
      $|\,\Gamma_5 (\lambda_5, \lambda_{12})\,|$ & 0.5903165 &
      $|\,\Gamma_5 (\lambda_{12}, \lambda_5)\,|$ & 0.5903165 \\
      $|\,\Xi_5 (\lambda_{11}, \lambda_5)\,|$ & 0.5903165 &
      $|\,\Xi_5 (\lambda_5, \lambda_{11})\,|$ & 0.5903165 \\
      $|\,\Xi_5 (\lambda_{12}, \lambda_6)\,|$ & 0.5903165 &
      $|\,\Xi_5 (\lambda_6, \lambda_{12})\,|$ & 0.5903165 \\
      $|\,\Gamma_5 (\lambda_{11},\lambda_6)\,|$ & 0.5903165 &
      $|\,\Gamma_5 (\lambda_6,\lambda_{11})\,|$ & 0.5903165 \\
      \hline\hline
    \end{tabular}
  }
\end{table}
The chiral Ward identity in Eq.~\eqref{eq:chiWard4} is exemplified in
Table~\ref{tab:WI}.
We find that the Ward identity holds valid up to our numerical
precision.

\section{Conclusion}
\label{sec:conc}
Using the chirality measurement and Ward identities, it is possible to
distinguish zero modes from non-zero modes reliably.
We will address this issue in more details in Ref.~\cite{prd:prep}.
%


\section*{Acknowledgement}
\begin{acknowledgement}
  We would like to express our sincere gratitude to Eduardo Follana
  for providing his code to us.
  We would like to express many thanks to Chulwoo Jung for providing
  the most updated CPS library to us.
  We would like to express our sincere gratitude to Jon Bailey and
  Stephen Sharpe for helpful discussion.
  The research of W.~Lee is supported by the Creative Research
  Initiatives Program (No.~2017013332) of the NRF grant funded by the
  Korean government (MEST).
  W.~Lee would like to acknowledge the support from the KISTI
  supercomputing center through the strategic support program for the
  supercomputing application research (No.~KSC-2015-G2-002).
  Computations were carried out in part on the DAVID clusters at
  Seoul National University.
\end{acknowledgement}

\bibliography{ref}

\end{document}